\documentclass[11pt,reprint,amsmath,amssymb,
 nofootinbib, aps,prl]{revtex4-2}
 
\usepackage{geometry}
\usepackage{amsmath,amsthm,amssymb, mathtools,mathrsfs}
\usepackage{physics}
\usepackage{amsfonts, latexsym, amssymb}
\usepackage{bbm, dsfont}
\usepackage{caption}
\usepackage{graphicx}
\usepackage{float}
\usepackage{url}
\usepackage{xcolor}
\usepackage{hyperref}
\usepackage{tikz}
\usetikzlibrary{decorations.pathmorphing,cd,decorations.markings,calc,fadings}
\hypersetup{
    colorlinks=true,
    linkcolor=blue,
    filecolor=magenta,      
    urlcolor=cyan,
}
\usepackage{enumerate}
\usepackage{bm}

\newcommand{\slashed}{{\bf\not}}

\newcommand{\eg}{{\it e.g.}}

\newcommand{\rrangle}{\rangle\!\rangle}

\newcommand{\nn}{\nonumber}

\newcommand{\be}{\begin{equation}} \newcommand{\ee}{\end{equation}}
\newcommand{\ben}{\begin{equation*}} \newcommand{\een}{\end{equation*}}

\newcommand{\bea}{\begin{equation} \begin{aligned}} \newcommand{\eea}{\end{aligned} \end{equation}}

\newcommand{\cB}{\mathcal{B}}

\newcommand{\calL}{\mathscr{L}}
\newcommand{\cM}{\mathcal{M}}

\newcommand{\bR}{\mathbb{R}}

\newcommand{\bZ}{\mathbb{Z}}

\newcommand{\fg}{\mathfrak{g}}

\newcommand{\ft}{\mathfrak{t}}

\def\repa{\raise4pt\hbox{$\square$}\mkern-14mu\raise-4pt\hbox{$\square$}}
\def\repab{\overline{\raise4pt\hbox{$\square$}\mkern-14mu\raise-4pt\hbox{$\square$}\mkern-1mu}}

\begin{document}

\title{
Taming Mass Gap with Anti-de-Sitter Space
}
\author{Christian Copetti$^{a}$, Lorenzo Di Pietro$^{b,c}$, Ziming Ji$^{c,d}$ and Shota Komatsu$^{e}$}
\affiliation{${}^{a}$Mathematical Institute, University of Oxford, Woodstock Road, Oxford, OX2 6GG, United Kingdom}
\affiliation{${}^{b}$Dipartimento di Fisica, Universit`a di Trieste, Strada Costiera 11, 34151 Trieste, Italy}
\affiliation{${}^{c}$INFN, Sezione di Trieste, Via Valerio 2, 34127 Trieste, Italy}
\affiliation{${}^{d}$SISSA, Via Bonomea 265, 34136 Trieste, Italy}
\affiliation{${}^{e}$Department of Theoretical Physics, CERN, 1211 Meyrin, Switzerland}
\email{christian.copetti@maths.ox.ac.uk}
\email{ldipietro@units.it}
\email{ziming.ji@sissa.it}
\email{shota.komatsu@cern.ch}

\date{\today}

\begin{abstract}
    Anti-de-Sitter space acts as an infra-red cut off for asymptotically free theories, allowing interpolation between a weakly-coupled small-sized regime and a strongly-coupled flat-space regime. We scrutinize the interpolation for theories in two dimensions from the perspective of boundary conformal theories. We show that the appearance of a singlet marginal operator destabilizes a gapless phase existing at a small size, triggering a boundary renormalization group flow to a gapped phase that smoothly connects to flat space.  We conjecture a similar mechanism for confinement in gauge theories.
\end{abstract}

\maketitle

\section{Introduction}
A striking aspect of quantum field theory (QFT) is the possibility that a classically marginal coupling becomes dimensionful at the quantum level, generating an energy scale $\Lambda$ from ``nothing''. This induces drastic changes in the low energy spectrum, particularly evident in examples like Yang-Mills (YM) theory in four dimensions. Despite classical expectations of massless excitations, the quantum theory exhibits a mass gap, a phenomenon yet to be rigorously proven for YM \cite{Jaffe:2000ne, Doug2004}.

Studying QFT in the Anti-de-Sitter space (AdS)  \cite{Callan:1989em} offers a new window into the dynamics through boundary conformal correlators \cite{Paulos:2016fap,Carmi:2018qzm,Komatsu:2020sag,Cordova:2022pbl,Antunes:2021abs,Levine:2023ywq,Meineri:2023mps,vanRees:2022zmr,vanRees:2023fcf,Giombi:2020rmc,Giombi:2021cnr,Giombi:2021uae}. The AdS radius ($L$) acts as a tunable parameter, enabling interpolation between a weakly-coupled regime ($\Lambda L\ll 1$) and a regime where the flat-space physics is recovered ($\Lambda L\gg 1$)  \cite{Paulos:2016fap,Carmi:2018qzm}. Previous work \cite{Aharony:2012jf} on YM in AdS$_4$ discussed an important consequence of the mass gap: Dirichlet boundary conditions (BC) at $\Lambda L \ll 1$ result in a family of boundary conformal theories, with a global symmetry mirroring the bulk gauge group and massless gluons in the bulk spectrum. These BC's cannot persist at $\Lambda L\gg 1$ due to the mass gap in flat space, hinting at a confinement transition, the precise mechanism of which remains elusive. Understanding the transition could aid in proving the mass gap applying  conformal bootstrap \cite{Poland:2018epd, Bissi:2022mrs, Hartman:2022zik} to the boundary correlators.

Motivated by these ideas, we study two-dimensional models in AdS$_2$ with a dynamically generated  gap; the $O(N)$ nonlinear $\sigma$-model and the $O(N)$ Gross-Neveu (GN) model, both of which have been studied since the discovery of asymptotic freedom as toy models for four-dimensional gauge theories. 

As in YM in four dimensions, we find families of BC's that exist only for small values of $\Lambda L$ and disappear at $\Lambda L\gg 1$.
  
In contrast to YM, in which the boundary theory at $\Lambda L\ll 1$ is characterized by the presence of a global symmetry,  these examples are characterized  by the presence of exactly marginal operators and an associated conformal manifold. This is the consequence of a BC that breaks a continuous global symmetry of the bulk theory, i.e.~the AdS analogue of the spontaneous symmetry breaking (SSB). These BC's must disappear at $\Lambda L \sim 1$ since the symmetry is unbroken in the flat-space vacuum. 

The transitions in these examples can be quantitatively analyzed using the large $N$ solvability of the theories. In both cases  the transition is due to a boundary operator, singlet under global symmetries of the boundary theory. It is irrelevant at weak coupling but becomes marginal at the transition point, triggering a RG instability.

We start by explaining general properties of these  transitions based on symmetry arguments and anomalies.  
We then move to 
explain the results of the large $N$ analysis.
\section{Tilt operators, boundary conformal manifolds and transitions}
Consider a QFT in AdS${}_{d+1}$ with a continuous symmetry $G$ and a boundary condition $|B \rrangle$. We let $\fg$ be the corresponding Lie algebra and $\mathfrak{t}^a$ its generators. The symmetry acts on $|B\rrangle$ through the topological~defects,
\be
U_\alpha = e^{i \alpha_g \int \star j} \, , \ \ \ \ \ \   j = \sum_a \, j_a \, \ft^a \, . \\
\ee
mapping $|B \rrangle$ to $|B_g \rrangle \equiv U_g \, | B \rrangle$. Generically $|B\rrangle$ is invariant only under a subgroup $G_\perp \subset G$, whose algebra will be denoted by $\fg_\perp$. The symmetry predicts the universal bulk-to-boundary OPE of the bulk conserved current,
\be
\star j_a(x,y) \sim \tau_a(x) + ...\qquad \ft^a \, \in \fg/\fg_\perp\,.
\ee
where $x$ and $y$ parametrize the boundary and the radial directions of AdS respectively, and $\tau_a$'s are called the {\it boundary tilt operators}. Generalizing the arguments of \cite{Herzog:2017xha}, we can show, using AdS isometries, that $
\tau_a$'s are exactly marginal. See the Supplemental Material. 
Turning them on, one obtains a boundary conformal manifold $\cM_{G/G_\perp}$ encoding the SSB of $G$ to $G_\perp$ in the bulk.
Unlike usual conformal manifolds encountered in CFTs, all points in $\cM_{G/G_\perp}$ are physically equivalent, as they are related by the action of the $G$ symmetry \cite{Drukker:2022pxk, Herzog:2023dop}. 

One might expect $\cM_{G/G_\perp}$ to exist at any  $\Lambda L$ as the symmetry $G$ does. This turns out to be incorrect.
We found that a marginal $G_\perp$-singlet operator appears in the boundary spectrum at $\Lambda L\sim 1$.
One can show under rather general assumptions that this  leads to the disappearance of the BC and triggers the RG flow \cite{Hogervorst:2021spa,Lauria:2023uca}.\footnote{A similar mechanism for the screening of conformal line defects was recently discussed in a series of works \cite{Cuomo:2022xgw,Aharony:2022ntz,Aharony:2023amq}.}
 This boundary transition generates  bulk mass gap, for which
we found two scenarios (see Figure \ref{fig: transtype}):
\begin{enumerate}
\item[$a)$] Continuous (``second-order") transition: $\cM_{G/G_\perp}$ shrinks to zero size, merging into a $G$-preserving BC. The BCFT data, as well as the bulk mass gap, change continuously across the transition point. This happens in the $O(N)$ $\sigma$-model.
\item[$b)$] Discontinuous (``first-order") transition: At \emph{finite} size, $\cM_{G/G_\perp}$ becomes unstable and undergoes a transition. In this case, the BCFT data jump discontinuously. It is sometimes possible to argue for a discontinuous transition based on mixed  anomaly, which forbids the merger of BC's with different symmetries. This happens in the GN model.
\end{enumerate}
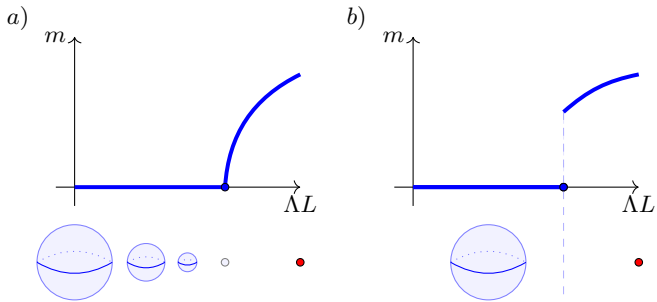
\begin{figure}

   \begin{tikzpicture}
   \node at (-0.75,2.25) {$a)$};
    \draw[->] (-0.25,0) -- (3,0) node[below] {$\Lambda L$};
    \draw[->] (0,-0.25) -- (0,2) node[left] {$m$};
    \draw[blue, line width=1.5] (0,0) -- (2,0) to[bend left=30] (3,1.5);
    \draw[fill=blue] (2,0) circle (0.05);
\begin{scope}[shift={(0,-1)}, scale=0.5]
\draw[blue, dotted] (-1,0) to[bend left=30] (1,0);
\draw[color=blue,fill=white!90!blue,opacity=0.5] (0,0) circle (1);
\draw[blue] (-1,0) to[bend right=30] (1,0);
\end{scope}     
\begin{scope}[shift={(0.95,-1)}, scale=0.25]
\draw[blue, dotted] (-1,0) to[bend left=30] (1,0);
\draw[color=blue,fill=white!90!blue,opacity=0.5] (0,0) circle (1);
\draw[blue] (-1,0) to[bend right=30] (1,0);
\end{scope}  
\begin{scope}[shift={(1.5,-1)}, scale=0.125]
\draw[blue, dotted] (-1,0) to[bend left=30] (1,0);
\draw[color=blue,fill=white!90!blue,opacity=0.5] (0,0) circle (1);
\draw[blue] (-1,0) to[bend right=30] (1,0);
\end{scope}  
\draw[fill=white!90!blue,opacity=0.5] (2,-1) circle (0.05);
\draw[fill=red] (3,-1) circle (0.05);

\begin{scope}[shift={(4.5,0)}]
\node at (-0.75,2.25) {$b)$};
     \draw[->] (-0.25,0) -- (3,0) node[below] {$\Lambda L$};
    \draw[->] (0,-0.25) -- (0,2) node[left] {$m$};
    \draw[blue, line width=1.5] (0,0) -- (2,0);
    \draw[white!90!blue, dashed] (2,0) -- (2,1);
    \draw[fill=blue] (2,0) circle (0.05);
    \draw[blue, line width=1.5] (0,0) -- (2,0);
    \draw[white!70!blue, dashed] (2,1) -- (2,-1.5);
    \draw[blue, line width=1.5] (2,1) to[bend left=15] (3,1.5);
    \draw[fill=blue] (2,0) circle (0.05);
    \begin{scope}[shift={(1,-1)}, scale=0.5]
\draw[blue, dotted] (-1,0) to[bend left=30] (1,0);
\draw[color=blue,fill=white!90!blue,opacity=0.5] (0,0) circle (1);
\draw[blue] (-1,0) to[bend right=30] (1,0);
\end{scope}     
\draw[fill=red] (3,-1) circle (0.05);
    \end{scope}
    \end{tikzpicture}
    \caption{Two scenarios for the phase transition. $a)$~Continuous transition.
    $b)$~Discontinuous transition.
    }
    \label{fig: transtype} 
\end{figure}

\section{Large $N$ examples}
We now study two large $N$ examples with a boundary phase transition and a dynamical bulk mass generation. The examples cover both scenarios described above.

\subsection{O(N) sigma model} 
Consider the 2d $O(N)$ $\sigma$ model, whose effective action at large $N$ is given by
\be
\mathscr{S} = \int \frac{1}{2} (\partial \phi)^2 + \sigma\left( \phi^2 - \frac{N}{g^2} \right) + N \text{tr} \log\left[ -\Box + 2 \sigma  \right] \, .
\ee
In flat space, the vacuum preserves $O(N)$ symmetry and $\phi^i$'s  acquire  mass $M \sim \Lambda = \mu e^{- \frac{2 \pi}{g^2}}$. By contrast, in AdS$_2$ the model has a weakly-coupled SSB phase with $N-1$ massless Goldstone fields  \cite{Carmi:2018qzm} and a VEV $\langle\phi^i\rangle = \sqrt{N}\Phi^i$, with $(\Phi)^2 = -\frac{1}{2\pi} \log(\Lambda L)$. This does not contradict the Coleman-Mermin-Wagner theorem \cite{mermin1966absence,coleman1973there} as AdS space comes with a natural IR regulator.

\paragraph{Boundary conditions.}Imposing the Dirichlet BC~$\phi^i |_{\partial AdS} = \Phi^i$ in the SSB phase, one obtains a  boundary theory with a conformal manifold $\cM_{O(N)/O(N-1)} \equiv \cM_{S^{N-1}}$.  Goldstone modes can be seen explicitly by expanding $\phi^i$ around the VEV $\Phi^i = \Phi \, n^i$:
\be \label{eq: phisplit}
\phi^i = (\sqrt{N}\Phi + \rho) n^i + \pi^i \, , \ \ \ \pi \cdot n^i = 0 \, .
\ee
The tilt operators $\tau^i = \partial_y \pi^i$ parametrize the tangent space $\text{T}S^{N-1}$ and arise from the pullback of the bulk $O(N)$ current\footnote{Concretely the $O(N)$ current is $j_\mu^{ij} = \phi^{\left[i \right.} \partial_\mu \phi^{\left. j \right]}$, using the small $y$ expansion $\phi^i = \Phi^i + y \partial_y \phi^i$ we find that $n_i j^{i \, j}_y = \Phi \partial_y \pi^j$.}. The presence of $\cM_{S^{N-1}}$ ensures that the bulk Goldstone modes remain massless. 

By contrast the $O(N)$-singlet BC~$\Phi^i = 0$ corresponds to the standard symmetry-preserving phase. It exists for $\Lambda L \geq \frac{1}{4}$ and all the way to the flat space limit $\Lambda L \gg 1$.

These two BC's merge at $\Lambda L=1$ where the radius of $\cM_{S^{N-1}}$ shrinks to zero. 
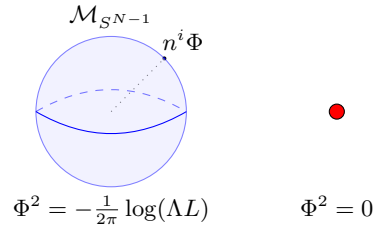
\begin{figure}
\begin{tikzpicture}
\draw[blue, dashed] (-1,0) to[bend left=30] (1,0);
\draw[dotted] (0,0) to (45:1);
\draw[fill=blue] (45:1) circle (0.02) ;
\node at (45:1.35) {$n^i \Phi$};
\draw[color=blue,fill=white!90!blue,opacity=0.5] (0,0) circle (1);
\draw[blue] (-1,0) to[bend right=30] (1,0);
\node[above] at (0,1) {$\cM_{S^{N-1}}$};
\node[below] at (0,-1) {$\Phi^2 = -\frac{1}{2\pi} \log(\Lambda L)$};
\begin{scope}[shift={(3,0)}]
\draw[fill=red] (0,0) circle (0.1);
\node[below] at (0,-1) {$\Phi^2 = 0$};
\end{scope}
\end{tikzpicture}
\caption{The boundary conformal manifolds for $O(N)$-breaking and $O(N)$-preserving boundary conditions in the $O(N)$ sigma model.}
\end{figure}
The merger is signalled by the appearance of a marginal boundary operator $\hat{\sigma}$ on $\cM_{S^{N-1}}$, the lightest scalar in the $\sigma$ bulk-to-boundary OPE. 
In addition to these two BC's, the model has Neumann boundary conditions, which exist only at small $\Lambda L$. Their physics is described in the Supplemental Material.

\begin{figure}
    \centering
    \includegraphics[scale=0.8]{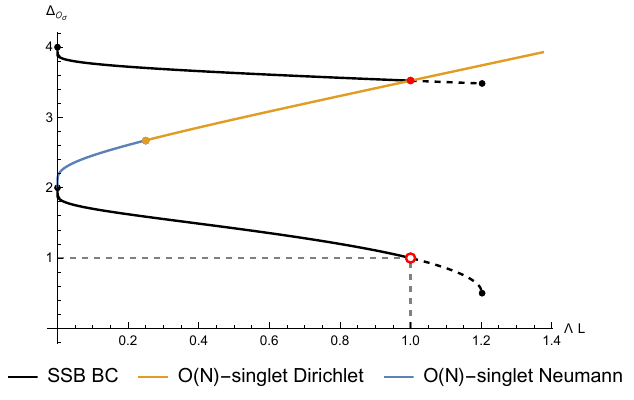}
    \caption{\label{fig: dimON}Dimensions of operators in the boundary spectrum of the bulk $\sigma$. Black lines are the lightest ($\hat{\sigma}$) and the second lightest ($\hat{\sigma}_2$) operators in the SSB BC~and the orange line is  the lightest operator ($\hat{\sigma}$) in the $O(N)$-singlet one. The blue line is for the Neumann BC.} 
\end{figure}

\paragraph{Details.}We now provide some details. We begin by studying the bulk phase structures following \cite{Carmi:2018qzm}. Minimizing the effective potential gives the gap equations
\bea
\Sigma \, \Phi^i &= 0 \, , \qquad
\Phi^2 - \frac{1}{g^2} + \text{tr} \left[ \frac{1}{- \Box + 2 \Sigma} \right] &= 0 \, .
\eea
A SSB solution requires $\Sigma=0$. Using the regulated AdS$_2$ propagator \cite{Carmi:2018qzm} we absorb the divergence of the trace  in the definition of the regulated coupling $1/g^2_{\rm reg}$. Introducing the dynamically generated scale $\Lambda = \mu e^{- 2 \pi / g^2_{\rm reg}}$ we find:
\be
\Phi^2 = - \frac{1}{2\pi} \log(\Lambda L) \, ,
\ee
$\Phi^2$ should be interpreted as the radius of $\cM_{S^{N-1}}$. The spectrum of the $\sigma$ bulk-to-boundary OPE is extracted from the poles of the AdS propagator 
\be\label{eq:sigsigON}
\langle \sigma(\textbf{x}_1) \, \sigma(\textbf{x}_2) \rangle = - \frac{1}{2}\int_{-\infty}^\infty d \nu \, \frac{1}{B(\nu) + \frac{2 \Phi^2}{\nu^2 + \frac{1}{4}}} \, \Omega_\nu(\textbf{x}_1, \, \textbf{x}_2) \, , 
\ee
where $B(\nu) = \frac{1}{2 \pi } \frac{\left( \psi(i \nu/2 + 3/4) + \psi(- i \nu/2 + 3/4) +\log(4) + 2\gamma \right)}{\nu^2 + \frac{1}{4}}$ is the 2d bubble function, see \cite{Carmi:2018qzm} or \ref{app: scalars} the Supplemental Material.
The dimensions of the lightest ($\hat{\sigma}$) and the second lightest  ($\hat{\sigma}_2$) operator are shown in Figure \ref{fig: dimON}. At $\Lambda L =1$,
$\hat{\sigma}$ hits marginality and decouples from the spectrum. The decoupling can be seen in the bulk-to-boundary OPE coefficient $b_{\hat{\sigma} \sigma}^2$, given by the residue of the pole of (\ref{eq:sigsigON}). Near the transition point $(\Phi=0)$ it reads
\be
b_{\hat{\sigma} \sigma}^2 = \frac{144}{\pi ^2}\Phi^2+O(\Phi^4) \, .
\ee
After the transition $b_{\hat{\sigma}\sigma}$ becomes imaginary, showing that the SSB BC becomes a complex BCFT \cite{Gorbenko:2018ncu}.

The $O(N)$-singlet BC~($\Phi^{i}=0$) can be analyzed similarly. The Breitenlohner-Freedman (BF) bound in two dimensions is $m^2L^2 \geq - \frac{1}{4}$. Since the mass of $\phi^i$ is $2 \Sigma$, this BC exists only for $
2\Sigma\geq -\frac{1}{4}$ that is $\Lambda L \geq \frac{1}{4}$.

At $\Lambda L=\frac{1}{4}$, the solution merges with the Neumann BC~discussed in the Supplemental Material, which continues to $\Lambda L\ll 1$, see Figure \ref{fig: dimON}.
If we instead increase $\Lambda L$, the dimension of $\hat{\sigma}$ intersects with that of $\hat{\sigma}_2$ of the SSB BC~at $\Lambda L=1$, leading to a continuous interpolation of the two BC's. Continuing further, it reproduces the known flat space results at $\Lambda L \gg 1$.
As $\hat{\sigma}$ is always irrelevant, the $O(N)$-singlet BC~is stable all the way to the strong coupling regime.

\subsection{Gross-Neveu model} 
The second example is the Gross-Neveu model, with the large $N$ effective action
\be
\mathscr{S} = \int \bar{\psi}_i \slashed{\nabla} \psi^i + \sigma \bar{\psi}_i \psi^i - \frac{N}{2g} \sigma^2 - N \, \text{tr} \log \left[ \slashed{\nabla} + \sigma \right]
\ee
 $\psi^i$ ($i=1, \, ... , \, N$) being  $N$ Dirac fermions. At finite $g$, in addition to $(-1)^{F}$, the model has a symmetry:
\be
G = O(2N)_V \times \bZ_2^{F_L}\, .
\ee
$\bZ_2^{F_L}$ is left-moving fermion parity acting by $\psi^{i}\to \gamma_3\psi^{i}$ and $\sigma \to - \sigma$. $G$ contains a $\bZ_4^A$ axial symmetry, generated by $\bZ_2^{F_L}$ together with a discrete $O(2N)_V$ rotation sending $\psi \to i \psi$. 
The axial symmetry thus acts as
\be
\psi^i \ \longrightarrow \ i \, \gamma_3 \, \psi^i \, , \ \ \ \sigma \ \longrightarrow \ - \sigma \, .
\ee
In flat space,  $\bZ_4^A$ is spontaneously broken to $\bZ_2^F$ by a $\sigma$ condensate, which gives a mass to the Dirac fermions \cite{gross1974dynamical}. 
\paragraph{Boundary conditions.}
This model also has two natural  BC's, which preserve either the vector $O(2N)_V$ or the axial $\bZ_4^A$.
The two cannot be preserved at the same time due to the mixed anomaly:
\be
I = \pi i \int A_A \cup c_1(F_V) \, . \label{eq: mixedan}
\ee

 $\bZ_4^A$-preserving BC's exist at weak coupling. They read, in Weyl components\footnote{$\psi=(\chi_L,\chi_R)^{t}$ and $\bar{\psi}=(\chi_R^{\ast},-\chi_L^{\ast})$}:
\be
|A; \, \eta \rrangle \, : \ \ \ (\chi_L^*)_i = e^{i \eta} (\chi_R)^i \, .
\ee
These preserve an $(\text{O}(N) \times \bZ_4^A)/\bZ_2^F$ symmetry\footnote{The $\text{O}(N)$ symmetry comes from an $U(N)^*$ symmetry of the free theory, which acts on the Weyl components as $\chi_L \ \to \ U \chi_L$ and $\chi_R \ \to \ U^* \chi_R$. The intersection of this group with the vector $O(2N)_V$ gives $O(N)$.}.
 The real scalar $\eta \, \in \, [0, \, 2 \pi)$ parametrizes an $S^{1}$ submanifold of $\cM_{O(2N)/O(N)}^A$ whose tilt operators include the pullback of the $U(1)_V$ current.
In this BC, $\sigma$ is not allowed to condense and the bulk fermions remain massless.

$O(2N)_V$ vector-preserving BC's are given by
\be
|V; \pm \rrangle \, : \ \ \ (\chi_L)^i = \pm (\chi_R)^i \, ,
\ee
These are the BC's associated to SSB of the axial symmetry. For each BC ($\pm$), there are two physically-distinct solutions to the gap equation. One of them (``Neumann-like") exists only at small $\Lambda L$ while the other (``Dirichlet-like") continues to the flat-space limit $\Lambda L\gg 1$. Here we focus on the Dirichlet-like $|V;\pm\rrangle_{D}$ leaving the other $|V;\pm\rrangle_{N}$ to the Supplemental Material. 
The two Dirichlet BC's $|V;\pm\rrangle_{D}$ are exchanged by the axial symmetry.

\begin{figure}
\includegraphics[scale=0.6]{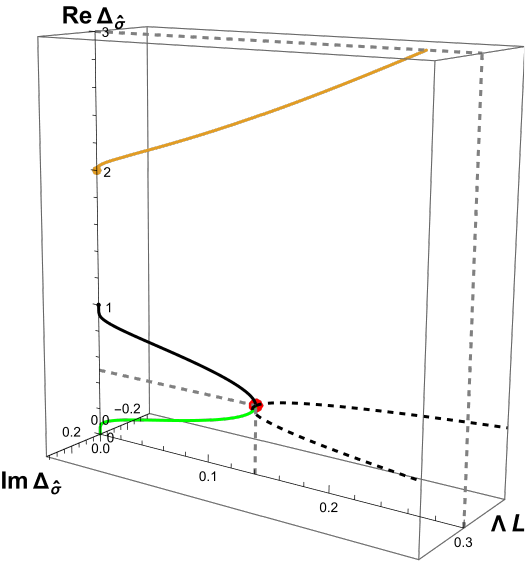}
    \caption{Dimensions of $\hat{\sigma}$ for $|V;\pm\rrangle_{D}$ (orange), $|A;\eta\rrangle$ (black), and $\widetilde{|A;\eta\rrangle}$ (green) in the GN model.}
    \label{fig: GNsigma}
\end{figure}

The dimensions of the lightest operator $\hat{\sigma}$ in the bulk $\sigma$ OPE in these BC's are given in Figure \ref{fig: GNsigma}. In addition, there is a ``shadow" of $\mathbb{Z}_4^{A}$-preserving BC, denoted by $\widetilde{|A;\eta\rrangle}$. At $\Lambda L\ll 1$, it arises from deforming $|A;\eta\rrangle$ by a double-trace operator $\mathcal{O}_{\sigma}=\hat{\sigma}^2$. As usual in the double-trace deformation at large $N$ \cite{Witten:2001ua, Gubser:2002zh, Gubser:2002vv}, the dimensions of $\hat{\sigma}$ in $|A;\eta\rrangle$ and $\widetilde{|A;\eta\rrangle}$ are related by $\Delta_{\hat{\sigma}}\to 1-\Delta_{\hat{\sigma}}$.

The BC's $|A;\eta\rrangle$ and $\widetilde{|A;\eta\rrangle}$ merge at $\Lambda L=\frac{e^{-\gamma}}{4}$. At the merger, the double-trace operator $\mathcal{O}_{\sigma}$  becomes marginal, triggering a RG flow whose endpoint we conjecture to be the vector-preserving BC, $|V,\pm\rrangle_{D}$.
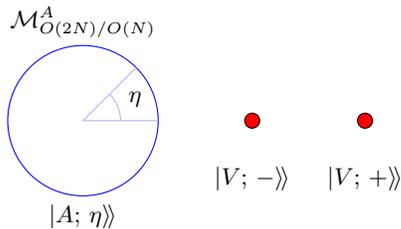
\begin{figure}
\begin{tikzpicture}
    \draw[color=blue] (0,0) circle (1);
    \node[above] at (0,1) {$\cM_{O(2N)/O(N)}^A$};
    \draw[blue!30!white] (1,0) -- (0,0) -- (45:1);
    \draw[blue!30!white] (0.5,0) arc (0:45:0.5 and 0.5); 
    \node at (22.5:0.75) {$\eta$};
    \node[below] at (0,-1) {$|A; \, \eta\rrangle$};
    \begin{scope}[shift={(3,0)}]
        \draw[fill=red] (-0.75,0) circle (0.1);
        \draw[fill=red] (0.75,0) circle (0.1);
        \node[below] at (-0.75,-0.5) {$|V; \, - \rrangle$};
        \node[below] at (0.75,-0.5) {$|V; \, +\rrangle$};
    \end{scope}
\end{tikzpicture}
    \caption{Maximally symmetric boundary conformal manifolds for the GN model.}
    \label{fig: boundaryGN}
\end{figure}
This transition is discontinuous and the BCFT data jump across the transition, as the axial and vector-preserving BCs cannot merge due to mixed  anomaly \eqref{eq: mixedan} \cite{Jensen:2017eof,Thorngren:2020yht}.  

\paragraph{BKT transition.}Abelian bosonization sheds new light on the transition. The bosonized theory is an $N$-component compact scalar $X^i$ with a 
potential for the T-dual coordinate $\tilde{X}$ \cite{Witten:1977xv, Ogilvie:1981ya}:
\be
V \sim g \sum_{i\neq j} \cos(2 \tilde{X}_i) \cos(2 \tilde{X}_j)  \, ,
\ee
The chiral symmetry is mapped to the $\bZ_4^w$ winding symmetry preserved by the potential. Since the fermion bilinear becomes an operator carrying two units of winding charge, its condensation correspond to proliferation of vortices {\it i.e.}~it describes an AdS analogue of the Berezinskii–Kosterlitz–Thouless (BKT) transition.

\paragraph{Details.}Let us provide some details. The gap equation reads:
\be\label{eq: FermionGap}
\frac{\Sigma}{g} + \text{tr} \left[ \frac{1}{\slashed{\nabla} + \Sigma} \right] = 0 \, .
\ee
In the Supplemental Material we show that $\tr[\frac{1}{\slashed{\nabla}}] = 0$ with the $|A; \, \eta\rrangle$ BC, by computing the propagator, see eq.~\eqref{eq:AxProp}. As a result the gap equation admits the axial symmetry preserving solution $\Sigma=0$.  

 This BC~is stable at $\Lambda L\ll 1$
\footnote{At weak coupling we must consider dimension-one bulk operators. The only scalar is the mass operator, which is forbidden by axial symmetry. The normal components of the currents are either trivial or protected. Parallel ones are odd under the reflection symmetry 
$\Tilde{R} = (-)^{F_L} \, R $
of the $|A; \, \eta \rrangle$ boundary condition, where $R$ acts by:
\be
R \, \psi^i(x,y) = \gamma^x \, \psi^i(-x, y) \, , \ \ \ R \, \bar{\psi}^i(x,y) = - \bar{\psi}^i(-x,y) \, \gamma^x \, .
\ee
}
and its boundary spectrum can be read off from the propagator of $\sigma$:
\be
N\langle \sigma(\textbf{x}_1) \, \sigma(\textbf{x}_2) \rangle = -\int_{-\infty}^{\infty} d \nu \frac{1}{g^{-1} - B_A(\nu)} \, \Omega_\nu(\textbf{x}_1, \, \textbf{x}_2) \, ,
\ee
where the bubble function $B_A(\nu)$ is computed in the Supplemental Material from the spectral transform of the square of the propagator $\tr[\frac{1}{\slashed{\nabla}^2}]$, see eq.~\eqref{eq:BAxial} for the result. This calculation involves a regularization that requires us to trade $g$ with the scale $\Lambda$.

As shown in Figure \ref{fig: GNsigma},
the dimension of  $\hat{\sigma}$ decreases monotonically from 1 as we increase $\Lambda L$. At $\Lambda L=\frac{e^{-\gamma}}{4}$, it hits the BF bound $\Delta=\frac{1}{2}$, making $\mathcal{O}_{\sigma}=\hat{\sigma}^2$ marginal at large $N$. We conjecture that this triggers the RG flow to $|V;\pm\rrangle_D$. In the proposed scenario, the operator $\hat{\sigma}^2$ which preserves the axial symmetry triggers a flow to a  BC which breaks it. This can happen in boundary conformal theories, as is  well-known in the RG flow from the special to the extraordinary BCFT's of the Ising model \cite{Cardy:1996xt}, triggered by a $\mathbb{Z}_2$ preserving boundary operator. 

The bubble function
 $B_V(\nu)$ for $|V;\pm\rrangle_D$ has been computed in \cite{Carmi:2018qzm}, which  we review in the Supplemental Material.
From this, one can check that $\hat{\sigma}$ in $|V;\pm\rrangle_{D}$ is always above marginality, ensuring its stability, see Figure \ref{fig: GNsigma}.

\section{Conjecture for Yang-Mills}Based on our findings, we now conjecture a mechanism for confinement of 
$SU(N)$ YM in AdS$_4$. At $\Lambda L\ll 1$, the Dirichlet BC leads to $SU(N)$ boundary conserved currents $J_{\mu}$. We first restate three scenarios, proposed in \cite{Aharony:2012jf},
for the disappearance of this BC at $\Lambda L\sim 1$, from the perspective of boundary conformal theories:
\begin{enumerate}
\item {\bf Higgs} (continuous): An operator $\mathcal{O}_{\rm adj}$ in the adjoint representation of $SU(N)$ hits marginality, leading to a multiplet recombination $\partial_\mu J^{\mu}\sim \mathcal{O}_{\rm adj}$ and giving anomalous dimension to $J_{\mu}$. The leading candidate for $\mathcal{O}_{\rm adj}$ is $\left.J_{\mu}J^{\mu}\right|_{\rm adj}$.
\item {\bf Decoupling} (continuous): The two-point function of $J_{\mu}$ vanishes and $J_{\mu}$'s decouple from the spectrum. %
\item {\bf Tachyon} (discontinuous): A scalar operator hits the BF bound, destabilizing the BC.
\end{enumerate}
Scenario 1 seems improbable as mentioned in \cite{Aharony:2012jf}, due to the difference of numbers of states between multi-gluons and glueballs. Scenario 2, sharing a decoupling feature with the $O(N)$ model, is plausible, but the post-decoupling BC remains to be clarified. Scenario 3 faces the issue that the operator becomes marginal\footnote{All scalar operators at $\Lambda L\ll 1$ are irrelevant.} before it hits the BF bound, triggering the RG flow earlier \cite{Lauria:2023uca}.

We propose an alternative:
\begin{itemize}
\item[4.] {\bf Marginality}: A scalar operator $\mathcal{O}_{\rm sing}$ 
 singlet under $SU(N)$ becomes marginal, triggering the RG flow to the Neumann BC. The leading candidate for $\mathcal{O}_{\rm sing}$  is ${\rm tr}(J_{\mu}J^{\mu})$.
\end{itemize}
The transition in this scenario could be either continuous or discontinuous. If continuous, Dirichlet and Neumann BC's smoothly merge, implying the decoupling of $J_{\mu}$\footnote{This is because there are no counterparts in the Neumann BC at $\Lambda L\sim 1$; e.g.~all gluon states are massive, having higher $\Delta$.}.  This can be seen as a refined version of Scenario 2.

However, we think that a discontinuous transition is more likely. This can be argued for example if we consider YM with gauge group   
 $SU(4)/\bZ_2$ or $\text{SO}(6)$. These have $\bZ_2^e \times \bZ_2^m$ one-form symmetry with mixed  anomaly
\be
I = \pi i \int B_e \, \cup \, \beta(B_m) \, . \label{eq: oneformanom}
\ee
Similar to the GN example,  Dirichlet and Neumann BC's preserve different symmetries ($\bZ_2^m$ and $\bZ_2^e$) and their merger is forbidden by the 't Hooft anomaly \eqref{eq: oneformanom}.

\section{Discussion}
We initiated the study of strong coupling effects in AdS through the lens of boundary conformal theories. 
The key signature is the appearance of singlet  marginal operators, leading to a transition from gapless to gapped phases.
 There are countless avenues for future explorations:
\begin{enumerate}
    \item We relied on large $N$ techniques to extract results valid at strong coupling. It is interesting to explore  alternatives \eg~the conformal bootstrap, semiclassics, and resurgence \cite{Dunne:2012ae}.
    \item In the $O(3)$ sigma model, the mass generation can be understood through the proliferation of instantons \cite{Berg:1979uq}. It is interesting to perform similar analysis in AdS and understand the interplay between semiclassical saddles and the stability of BC's.
    \item Also  interesting is to test our conjecture on confinement in YM. In particular, if the transition is discontinuous, one can use the current two-point function $\langle J J\rangle$ to parametrize the AdS radius, and  see if solutions to the conformal bootstrap stop existing at  nonzero $\langle JJ\rangle$. In practice,  more inputs (than the existence of $SU(N)$ currents) might be needed and  a key  question is to figure them out.
\end{enumerate}
\paragraph{Acknowledgements} We thank Francesco Benini, Simone Giombi, Zohar Komargodski, Edoardo Lauria, Marco Meineri, Kyriakos Papadodimas, Balt van Rees, and Yunqin Zheng for illuminating discussions. CC is supported by STFC grant  ST/X000761/1. ZJ is supported by the ERC-COG grant NP-QFT No. 864583 “Non-perturbative dynamics of quantum fields: from new deconfined phases of matter to quantum black holes” and by the MIUR-SIR grant RBSI1471GJ. LD and ZJ acknowledge support from the INFN “Iniziativa Specifica ST\&FI”.

\bibliographystyle{utphys}
\bibliography{AdSflows_draft_v3}


\pagebreak
\clearpage
\widetext\begin{center}
\textbf{\large Supplemental Materials: }
\end{center}

\section{Scalars in $\text{AdS}_{d+1}$}\label{app: scalars}
In order to treat scalar fields in AdS let us introduce the harmonic function $\Omega_\nu(\textbf{x},\textbf{y})$ which is an eigenfunction of the AdS Laplacian:
\be
-\square_{\bold{x}} \, \Omega_\nu(\textbf{x},\textbf{y}) = \frac{1}{L^2}(\nu^2 + \frac{d^2}{4}) \, \Omega_\nu(\textbf{x},\textbf{y}) \, ,
\ee
a scalar two-point function $F(\textbf{x}, \textbf{y})$ can be expanded in harmonics
\be
F(\textbf{x},\textbf{y}) = \int_{-\infty}^\infty d\nu \, F(\nu) \, \Omega_\nu(\textbf{x}, \textbf{y}) \, .
\ee
The spectral decomposition has the further property that convolutions in position space become standard products in $\nu$ space. For example the massive free scalar propagator $G_\Delta(\textbf{x},\textbf{y})$ becomes:
\be
G_\Delta(\textbf{x},\textbf{y}) = \frac{1}{L^{d-1}} \int_{-\infty}^{\infty} d \nu \frac{1}{\nu^2 + (\Delta - \frac{d}{2})^2} \, \Omega_\nu(\textbf{x},\textbf{y}) \, .
\ee
We now turn our attention to the $O(N)$ sigma model. The propagators in both phases were determined in \cite{Carmi:2018qzm}. One starts by expanding the $O(N)$ Lagrangian around the vacuum configuration. In the symmetry-preserving phase, this quadratic Lagrangian reads:
\be
\calL^{(2)} = \frac{1}{2} (\partial \phi)^2 + \Sigma \, \phi^2 + \sigma \, \left(\frac{1}{-\Box + 2 \Sigma} \right)^2 \, \sigma  \, .
\ee
Defining the bubble function
\be
B(\textbf{x},\textbf{y}) = \left[ \frac{1}{-\Box + 2 \Sigma} \right]^2(\textbf{x}, \, \textbf{y}) \, ,
\ee
the $\sigma$ propagator reads:
\be
\langle \sigma(\textbf{x} ) \, \sigma(\textbf{y}) \rangle = B^{-1}(\textbf{x}, \, \textbf{y}) \, .
\ee
The bubble function has been bootstrapped in \cite{Carmi:2018qzm} to be, in the spectral representation:
\be
\begin{aligned}
 &\tilde{B}(\nu)=\frac{\Gamma_{\Delta}\Gamma_{\Delta-\frac{d}{2}+\frac{1}{2}}\Gamma_{2\Delta-\frac{d}{2}}}{4(4\pi)^{\frac{d}{2}}}\\
 &\times \left(\Gamma_{\Delta-\frac{d+2i\nu}{4}}\,{}_5\tilde{F}_{4}\left[\begin{array}{c}\{\frac{d}{2},\Delta,\Delta-\frac{d}{2}+\frac{1}{2},\Delta-\frac{d+2i\nu}{4},2\Delta-\frac{d}{2}\}\\\{\Delta+\frac{1}{2},\Delta-\frac{d}{2}+1,\Delta-\frac{d+2i\nu}{4}+1,2\Delta-d+1\}\end{array};1\right]\right.\\
 &\left.\quad+ \Gamma_{\Delta-\frac{d-2i\nu}{4}}\,{}_5\tilde{F}_{4}\left[\begin{array}{c}\{\frac{d}{2},\Delta,\Delta-\frac{d}{2}+\frac{1}{2},\Delta-\frac{d-2i\nu}{4},2\Delta-\frac{d}{2}\}\\\{\Delta+\frac{1}{2},\Delta-\frac{d}{2}+1,\Delta-\frac{d-2i\nu}{4}+1,2\Delta-d+1\}\end{array};1\right]\right)\,,
 \end{aligned}
\ee
with $\Delta = \frac{d}{2} + \sqrt{\frac{d^2}{4} + 2 \Sigma}$.
In the symmetry breaking phase the situation is slightly different. Using \eqref{eq: phisplit} the quadratic Lagrangian reads:
\be
\calL^{(2)} = \frac{1}{2} (\partial \pi)^2 + \frac{1}{2} (\partial \rho)^2 + 2 \Phi \rho \, \sigma + \sigma \, \left( \frac{1}{-\Box} \right)^2 \, \sigma
\ee
The propagator of $\rho$ and $\sigma$ is given, in the spectral representation, by the inverse of the matrix:
\be
\tilde{K}(\nu) = \begin{pmatrix}
-\tilde{B}(\nu)\big|_{\Sigma=0}     &  2\Phi\\
2\Phi & \nu^2 + \frac{d^2}{4} 
\end{pmatrix}
\ee
which immediately gives the expression in the main text after taking the first diagonal entry.

We will also need the following expansion for the dimensionally-regulated inverse propagator at $d=1 + \epsilon$:
\be
\begin{aligned} \label{eq: invpropON}
&\text{tr} \left[ \frac{1}{-\Box + 2 \Sigma} \right] = - \frac{1}{2\pi} \left[ \psi\left(\frac{1}{2} +\sqrt{\frac{1}{4} + 2 \Sigma}\right) + \gamma \right] \\ 
&+\frac{1}{2\pi}\left(- \frac{1}{\epsilon} + \gamma + \log(L \mu) + 2 \log(4\pi)\right) \, ,
&
\end{aligned}
\ee
The second line of equation \eqref{eq: invpropON} can be absorbed into the definition of the regularized coupling $1/g^2_{\text{reg}}$ we use in the main text.

\subsection{The Neumann boundary condition}
Let us give details about the physics with the Neumann BC's. The Neumann propagator  differs from the Dirichlet one as we must pick up the residue in the lower-half plane instead of the one in the upper-half plane:
\be
G_{\Delta_-}(\textbf{x}, \, \textbf{y}) = \frac{1}{L^{d-1}} \int_{\bR - C_u + C_l} \frac{1}{\nu^2 + (\Delta-\frac{d}{2})^2} \, \Omega_\nu(\textbf{x}, \textbf{y}) \, .
\ee
This affects the form of the gap equation, which for $d=1$ reads
\be
\Phi^2 + \frac{1}{2\pi} \log(\Lambda L)  - \frac{1}{2\pi} \left[ \psi\left(\frac{1}{2} +\sqrt{\frac{1}{4} + 2 \Sigma}\right) + \gamma \right] + \frac{1}{2} \tan\left(\pi \sqrt{\frac{1}{4} + 2 \Sigma} \right) = 0 \, ,
\ee
where the last term on the left hand side comes from the change of the contour. As the new term diverges as $\Sigma \to 0$, there are no symmetry breaking solutions with Neumann BC's. For the symmetry preserving ones, we set $\Phi=0$ and solve for $\Sigma$. Recall that in AdS$_2$, a scalar field can be quantized with Neumann boundary conditions only if its mass satisfies $-\frac{1}{4}\leq M^2L^2 \leq 0$. Since the mass squared of the $O(N)$ fields $\phi_i$ is $2\Sigma$, we need to impose
\be
- \frac{1}{4}\leq 2\Sigma\leq 0\quad \Rightarrow \quad 0\leq\Lambda L\leq \frac{1}{4}  
\ee
This shows 1.~the Neumann BC is incompatible with the flat spacetime limit, 2.~as we increase the AdS radius from zero, the mass square of $\phi_i$ descends from $1$ hitting the BF bound at $\Lambda L =\frac{1}{4}$, where it smoothly merges with the Dirichlet BC. At the merger point, two BC's become indistinguishable.  See Figure \ref{fig: dimON}. Different phases and their transitions can be summarized below:
\bea
\begin{tikzpicture}
\draw[->] (0,0) node[below] {$\Lambda L = 0$} -- (5,0) node[below] {$\Lambda L$};
\draw[dashed,->] (3,1.5) to (3,2.25); \draw[dashed] (3,2.25) to (3,3);
\draw[dashed,->] (1,4.5) -- (1,3.75); \draw[dashed] (1,3.75) -- (1,3);
\draw[black, line width=1]  (0,1.5) node[left] {D SSB} -- (3,1.5); \draw[red, fill=white] (3,1.5) circle (0.1) node[below] {$\Lambda L = 1$};
\draw[orange, line width=1] (1,3) node[left] {D $\Sigma \neq 0$} node[below, black] {$\Lambda L = \frac{1}{4}$} -- (5,3); \draw[fill=orange] (1,3) circle (0.1); \draw[orange, line width=1, dotted] (5,3) -- (6,3);
\draw[blue, line width=1] (0,4.5) node[left] {N} -- (1,4.5) node[above, black] {$M^2 L^2 = -\frac{1}{4}$}; 
\draw[fill=orange] (1,4.5) circle (0.1);
\end{tikzpicture}
\eea

\section{Fermions in $\text{AdS}_{d+1}$}

Let us now discuss the free massless fermion propagator with the $|A; \, \eta \rrangle$ boundary condition. We focus on $d=2$ for simplicity. We confomally map AdS$_2$ to the upper-half plane. Given a generic boundary condition:
\be
\varphi^a_L = M^{a}{}_b \, \varphi^b_R \, ,
\ee
on the Weyl-Majorana modes the method of images \cite{francesco2012conformal} gives the UHP correlators:
\bea
&\langle \varphi_L^a(z_1) \, \varphi_L^b (z_2) \rangle_{\cB} = \frac{\delta^{ab}}{z_{12}} \, , \ \ \  &&\langle \varphi_R^a(\bar{z}_1) \, \varphi_R^b (\bar{z}_2) \rangle_{\cB} = \frac{\delta^{ab}}{\bar{z}_{12}} \, , \\
&\langle \varphi_L^a(z_1) \, \varphi_R^b (\bar{z}_2) \rangle_{\cB} = \frac{M^{ba}}{(z_1 - z_2^*)} \, , \ \ \  &&\langle \varphi_R^a(\bar{z}_1) \, \varphi_L^b (z_2) \rangle_\cB = \frac{M^{ab}}{z_1^* - z_2} \, .
\eea
Going back to AdS$_2$ by the Weyl rescaling we find:
\bea
&\langle \varphi_L^a(z_1) \, \varphi_L^b (z_2) \rangle_{\cB}^{AdS} = \frac{z_{12}^*}{\sqrt{y_1 y_2}} \frac{\delta^{ab} \ }{\zeta} \, , \ \ \  &&\langle \varphi_R^a(\bar{z}_1) \, \varphi_R^b (\bar{z}_2) \rangle_{\cB}^{AdS} =  \frac{z_{12}}{\sqrt{y_1 y_2}} \frac{\delta^{ab}}{\zeta} \, , \\
&\langle \varphi_L^a(z_1) \, \varphi_R^b (\bar{z}_2) \rangle_{\cB}^{AdS} =  \frac{z_1^*-z_2}{\sqrt{y_1 y_2}} \frac{M^{ba}}{\zeta +4} \, , \ \ \  &&\langle \varphi_R^a(\bar{z}_1) \, \varphi_L^b (z_2) \rangle_\cB^{AdS} = \frac{z_1-z_2^*}{\sqrt{y_1 y_2}} \frac{M^{ab}}{\zeta + 4} \, ,
\eea
with $\zeta = \frac{\textbf{x}_{12}^2}{y_1 y_2}$ the AdS chordal distance. Using the Weyl basis $\chi^i = \varphi^{2i} + i \varphi^{2i +1}$ and the explicit formula for the $|A; \, \eta \rrangle$ boundary condition:
\be
M = \begin{bmatrix}
 \begin{array}{cc}
     \cos(\eta) & \sin(\eta) \\ \sin(\eta) & - \cos(\eta)
 \end{array}  &  & 0 \\
 ... & \ddots & ...  \\
  0 & ... &  \begin{array}{cc}
     \cos(\eta) & \sin(\eta) \\ \sin(\eta) & - \cos(\eta)
 \end{array}
\end{bmatrix}
\ee
one obtains
\be\label{eq:AxProp}
\langle \bar{\psi}^i(\textbf{x}_1) \, \psi^j(\textbf{x}_2) \rangle_A = - \frac{\Gamma((d+1)/2)}{2 \pi^{(d+1)/2}} \frac{\gamma_a \, \textbf{x}_{12}^a}{\sqrt{y_1 y_2}} \frac{\delta^{ij}}{\zeta^{(d+1)/2}}~.
\ee
Using the notation $\frac{1}{\slashed{\nabla}}$ for this propagator, it follows immediately that with this boundary condition $\tr\left[\frac{1}{\slashed{\nabla}}\right]=0$ as we claimed in the main text, simply because of the tracelessness of the $\gamma$ matrix.

The bubble function $B_A(\nu)$ is defined as the spectral transform of
\be
\text{Tr}_\gamma \left[\frac{1}{\slashed{\nabla}}\right]^2 = \frac{2^{\frac{d-3}{2}}}{\pi^{\frac{d+1}{2}}} \, \frac{\Gamma(\frac{d+1}{2})^2 \Gamma(\frac{1-d}{2}) }{\Gamma(d)} \, \frac{1}{\zeta^d} \, .
\ee
where $d$ is kept generic here as a UV regulator.We compute it using the general result for the spectral transform of a power of the chordal distance \cite{Hogervorst:2017sfd, Carmi:2018qzm}
\be
\widehat{\zeta^{-p}}(\nu) = (4 \pi)^{\frac{d+1}{2}} \, \frac{\Gamma(\frac{d+1}{2} - p)\Gamma(-\frac{d}{2} + p \pm i \nu)}{\Gamma(p) \Gamma(\frac{1}{2} \pm i \nu)} \, . 
\ee
Here we adopted the notation $\Gamma(a\pm b)\equiv \Gamma(a+b)\Gamma(a-b)$. The result is
\begin{align}
\begin{split}
B_A(\nu) & = \frac{1}{2^{(d+1)/2}\pi^{(d+1)/2}} \frac{\Gamma(\frac{d+1}{2})^2 \Gamma(\frac{1-d}{2})}{\Gamma(d)} \ \frac{\Gamma(\frac{d}{2} \pm i \nu)}{\Gamma(\frac{1}{2} \pm i \nu)}~.
\end{split}
\end{align}
Finally, to obtain the result for AdS$_2$, we plug $d=1+\epsilon$ and expand for $\epsilon\to 0$ and reinstate the AdS length $L$. The result is
\be\label{eq:BAxial}
\begin{aligned}
&B_A(\nu) \sim -\frac{1}{2 \pi}\left( \frac{2}{\epsilon} + \gamma - 2 \log(L \mu) + \log(2\pi) \right) \\
&- \frac{1}{2\pi}\left( \psi\left(\frac{1}{2} + i \nu\right) + \psi\left(\frac{1}{2} - i \nu\right) \right) \, .
\end{aligned} 
\ee
 We absorb the
 divergent and constant terms in the first line into the definition of the regulated coupling $g_{\text{reg}}$, and introduce the dynamically generated scale $\Lambda = \mu e^{-\frac{\pi}{g_{\text{reg}}}}$.

\subsection{The massive boundary conditions}
Let us also review the standard $U(N)_V$-preserving boundary conditions which are relevant for the study of the $\bZ_4^A$ symmetry broken phase. Near the AdS boundary a Dirac fermion with mass $\Sigma$ has the asymptotic expansion
\begin{equation}
\psi^i(y, x) \underset{y\to 0}{\longrightarrow}  y^{\Delta_+}(\psi_+^i(\vec{x})+\mathcal{O}(y) ) +  y^{\Delta_-}(\psi_-^i(\vec{x})+\mathcal{O}(y) )~,
\end{equation}
where we define $\Delta_{\pm} = \frac{d}{2}\pm \Sigma$ and the modes $\psi_\pm^i(\vec{x})$ satisfy
\begin{equation}
\gamma_\perp \psi_\pm^i = \pm \psi_\pm^i~.
\end{equation}
The two types of $U(N)_V$ preserving boundary conditions correspond to setting $\psi_-$ or $\psi_+$ to $0$, which we call $|V; + \rrangle$ and $|V; - \rrangle$ boundary condition, respectively. The axial symmetry swaps them and changes the sign of the mass gap at the same time. We also call $|V; + \rrangle$ with positive mass and $|V; - \rrangle$ with negative mass Dirichlet-like because $\Delta$ is larger than $\frac{d}{2}$, and $|V; + \rrangle$ with negative mass and $|V; - \rrangle$ with positive mass Neumann-like because $\Delta$ is smaller than $\frac{d}{2}$.
The gap equation \eqref{eq: FermionGap} is the same for the $+$ and $-$ boundary conditions, and is solved by first evaluating the spectral integral 
\begin{equation}
\begin{aligned}
&\text{tr} \left[ \frac{1}{\slashed{\nabla} + \Sigma} \right]=\\
&c_{d+1}\int_{-\infty}^{\infty} d\nu \left[\frac{1}{\nu^2+(\Sigma-\tfrac{1}{2})^2}-\frac{1}{\nu^2+(\Sigma+\tfrac{1}{2})^2}\right]\frac{\Gamma(\tfrac{d}{2})\Gamma(1+\tfrac{d}{2}\pm i\nu)}{4 \pi^{\frac{d}{2}+1}(d+1)\Gamma(d)\Gamma(\pm i\nu)}\,,
\end{aligned}
\end{equation}
where $c_{d+1}$ is the number of components of a Dirac spinor in $d + 1$ dimensions. Using dimensional regularization
\begin{equation}
\begin{aligned}
&\text{tr} \left[ \frac{1}{\slashed{\nabla} +  \Sigma} \right]_\text{dimreg}=\\
&c_{d+1}\frac{\pi ^{-\frac{d}{2}-1} \sec \left(\frac{\pi  d}{2}\right) \Gamma \left(\frac{d}{2}\right) }{4 (d+1) \Gamma (d)}\Big(\sin \left(\frac{\pi  \left(d-\left| \Sigma -1\right| \right)}{2} \right) \Gamma_{\frac{d}{2}- \left| \frac{\Sigma}{2} -\frac{1}{2}\right| +1} \Gamma_{\frac{d}{2}+\left| \frac{\Sigma}{2} -\frac{1}{2}\right| +1}\\
&\qquad\qquad\qquad\qquad\qquad\qquad -\sin \left(\frac{\pi  \left(d-\left| \Sigma +1\right| \right)}{2} \right) \Gamma_{\frac{d}{2}-\left| \frac{\Sigma}{2} +\frac{1}{2}\right| +1} \Gamma_{\frac{d}{2}+\left| \frac{\Sigma}{2} +\frac{1}{2}\right| +1}\Big)\,.
\end{aligned}
\end{equation}
Setting $d=1 + \epsilon$ this is
\begin{equation}
\begin{aligned}
\text{tr} \left[ \frac{1}{\slashed{\nabla} +  \Sigma} \right]_+=-\frac{\Sigma }{\pi  \epsilon}+\frac{\Sigma  \left(-2\psi(|\Sigma|)
-\gamma +\log (4 \pi )\right)-\text{sgn}(\Sigma)}{2\pi }+O\left(\epsilon\right)\,,
\end{aligned}
\end{equation}
Absorbing the divergent piece and some constants into $g_\text{reg}$ and using $\Lambda=\mu e^{-\frac{\pi}{g_\text{reg}}}$ and reinstating the AdS length $L$, the gap equation is 
\begin{equation}
    \frac{\Sigma}{\pi}\log(\Lambda L)-\frac{2\Sigma\psi(|\Sigma|)+\text{sgn}(\Sigma)}{2\pi}=0
\end{equation}
Then by solving the gap equation one can see that the mass gap is zero at $\Lambda L=0$ and monotonically increases as $\Lambda L$ increases and reaches $\Lambda$ in the flat space limit. The Dirichlet-like boundary conditions continue to exist at any $\Lambda L$ while the Neumann-like boundary conditions are only unitary at weak coupling. Thus from now on and in the main text, we only consider the Dirichlet-like boundary conditions.

We know that the vector-preserving boundary condition does not degenerate into the axial-preserving boundary condition even in the massless (free) limit by anomaly arguments. Here we comment on the leading spectrum of the Dirichlet-like boundary conditions at finite $\Lambda L$. The bubble function, in this case, has been computed in \cite{Carmi:2018qzm}:
\begin{align}
& \mathcal{B}_{\text{AdS}_2} (\Delta,\nu)=\frac{i\nu\Gamma_{\frac{1}{4}+\Delta}\Gamma_{\frac{1}{2}+\Delta}^2\Gamma_{\frac{1}{2}+2\Delta}}{8\sqrt{\pi}}\times\\
 &\left(\Gamma_{\Delta+\frac{1}{4}+\frac{i\nu}{2}}\,{}_6\tilde{F}_{5}\left[\begin{array}{c}\{\frac{3}{2},\frac{1}{4}+\Delta,\frac{1}{2}+\Delta,\frac{1}{2}+\Delta,\frac{1}{2}+2\Delta,\frac{1}{4}+\Delta+\frac{i\nu}{2}\}\\\{2\Delta,1+\Delta,1+\Delta,\frac{5}{4}+\Delta,\frac{5}{4}+\Delta+\frac{i\nu}{2}\}\end{array};1\right]-(\nu \leftrightarrow -\nu)\right)~,\nn
 \end{align}
with some constant shift, which we omit here. The dimension $\Delta$ is determined by the solution to the gap equation at finite $\Lambda L$. The final inverse two-point function in the spectral representation is \cite{Carmi:2018qzm}:
 \begin{align}
 (F_{\delta\sigma\delta\sigma}(\nu))^{-1}=\mathcal{B}_{\text{AdS}_2} (\Delta,\nu)-\mathcal{B}_{\text{AdS}_2} (\Delta,\tfrac{i}{2})+\frac{(\Delta-\tfrac{1}{2})\psi^{(1)}(\Delta-\tfrac{1}{2})}{\pi}-\frac{1}{\pi (2\Delta-1)}\,.
 \end{align}
We obtain the dimension of the lightest operators in the $\sigma$ bulk-to-boundary OPE by looking at the poles of the two-point function, which is plotted in figure \ref{fig: GNsigma}. In the free limit $\Lambda L=0$, the lightest operator in the $+$ boundary condition has dimension $2$ while the lightest operator in the axial-preserving boundary condition has dimension $1$. This is a direct consequence of the symmetries \cite{Carmi:2018qzm}. Note that at all $\Lambda L$, there is no operator becoming marginal in the Dirichlet-like boundary conditions.

\section{Tilt operator from current two-point function}

In this appendix we show that for a QFT in AdS$_{d+1}$, whenever a bulk continuous global symmetry is broken by the boundary condition, there is an exactly marginal operator on the boundary, the tilt operator. This is a generalization to massive theories in AdS of the well-known result in boundary conformal field theory \cite{Bray_1977,  Herzog:2017xha, Cuomo:2021cnb, Padayasi:2021sik}. In particular \cite{Herzog:2017xha}  studied the bulk two-point function of a conserved current in BCFT, and showed that whenever a scalar operator appears in the boundary OPE decomposition, it must have dimension $d$ and therefore be marginal. We now show that this argument generalizes to massive theories in AdS$_{d+1}$, using the isometries and conservation to constrain the possible form of the two-point function.\footnote{We thank Balt van Rees for suggesting that such an argument should exist.} In the following we will keep $d$ generic and distinguish between spin 1 and spin 0 operators appearing in the boundary OPE of the current. This terminology is appropriate for $d\geq 2$ where the boundary spin is well-defined, but in $d=1$ we can still distinguish between the two sets of operators using the discrete reflection symmetry on the boundary.

The two-point function of a spin 1 operator in AdS$_{d+1}$ admits the following integral representation \cite{Costa:2014kfa, Loparco:2023rug, Ankur:2023lum}
\begin{align}\label{eq:spinonedec}
\begin{split}
F(\bold{x},\bold{y})_{MN} & = \int_{-\infty}^{+\infty} d\nu~F^{\perp}(\nu)\Omega^{(1)}_{\nu\,MN}(\bold{x},\bold{y})+ \nabla^{\bold{x}}_M \nabla^{\bold{y}}_N \,\int_{-\infty}^{+\infty} d\nu~F^{L}(\nu)\Omega_\nu(\bold{x},\bold{y})~.
\end{split}
\end{align}
Here $M$ and $N$ are indices running in the tangent space at $\bold{x}$ and $\bold{y}$, respectively. The function $\Omega^{(1)}_{\nu\,MN}(\bold{x},\bold{y})$ is the spin one harmonic function in AdS. It satisfies
\begin{align}
\begin{split}
-\square_{\bold{x}} \Omega^{(1)}_{\nu\, MN}(\bold{x},\bold{y}) & = \frac{1}{L^2}\left(\nu^2 + \tfrac{d^2}{4}+1\right) \Omega^{(1)}_{\nu\, MN}(\bold{x},\bold{y})~,\\
\nabla_{\bold{x}}^M \Omega^{(1)}_{\nu\, MN}(\bold{x},\bold{y}) & = 0~,
\end{split}
\end{align}
and is symmetric under simultaneous exchange of $\bold{x}$ with $\bold{y}$ and $M$ with $N$. Given that $\Omega^{(1)}$ is divergence-free, we can think of the functions $F^\perp(\nu)$ and $F^L(\nu)$ as the transverse and the longitudinal parts of the two-point function. It follows that in the case of a conserved current we must have $F^L(\nu)=0$.\footnote{This assumes that the two-point function is conserved even at coincident points. While this is certainly the case for $d+1>2$, for $d+1=2$ contact terms in the divergence of the two-point function of the current are admissible and signal the existence of a quadratic 't Hooft anomaly. When these contact terms exist, we must have $F^L(\nu) \propto (\nu^2 + \frac{d^2}{4})^{-1}$ in order to reproduce them, where the proportionality constant is determined by the coefficient of the contact term. The corresponding pole of $F^L(\nu)$ in the complex $\nu$ plane is precisely at the location associated to a scalar operator of dimension $d$, therefore we see that even allowing contact terms does not change the main conclusion.}

Using the expression 
\begin{equation}
\Omega^{(1)}_{\nu\, MN}(\bold{x},\bold{y}) = \frac{i\nu}{2\pi}\left(G^{(1)}_{\frac{d}{2}+i\nu,MN}(\bold{x},\bold{y})- G^{(1)}_{\frac{d}{2}-i\nu, MN}(\bold{x},\bold{y})\right)~,
\end{equation}
where $G^{(1)}_{\Delta,MN}$ is the Proca AdS propagator for a spin one particle of mass $M^2=\frac{1}{L^2}(\Delta-1)(\Delta -d +1)$, we can separate the function $\Omega^{(1)}_{\nu\, MN}$ in two terms that allow to close the contour in the $\nu$ integral, in the lower ($G^{(1)}_{\frac{d}{2}+i\nu,MN}$) or upper ($G^{(1)}_{\frac{d}{2}-i\nu,MN}$) half plane. Picking up the poles of the transverse function $F^\perp(\nu)$, we obtain a decomposition of the two-point function in discrete contributions proportional to $G^{(1)}_{\Delta,MN}$, associated to the exchange of a boundary operator of spin 1 and dimension $\Delta$. This decomposition is precisely the boundary OPE decomposition of the two-point function. Therefore, the poles of $F^\perp(\nu)$ tell us about the spectrum of spin one operators appearing in this OPE.

From this discussion, it seems like there is no room for possible contributions of spin 0 operators to the two-point function of a conserved current. However this overlooks the fact that the function $G^{(1)}_{\frac{d}{2}\pm i\nu,MN}$ has itself a pole as a function of $\nu$, namely
\begin{equation}
\pm \frac{i\nu}{2\pi}G^{(1)}_{\frac{d}{2}\pm i\nu,MN}(\bold{x},\bold{y}) \underset{\nu \to \mp i( \frac{d}{2}-1)}{\sim} \pm\frac{i}{4\pi} \frac{1}{\nu \pm i(\frac{d}{2}-1)}\nabla^{\bold{x}}_M \nabla^{\bold{y}}_N G_d(\bold{x},\bold{y}) + \mathcal{O}((\nu \pm i(\tfrac{d}{2}-1))^0)~.
\end{equation}
Here, using the notation introduced above, $G_d$ denotes the AdS propagator for a massless scalar, associated to a boundary scalar operator of dimension $d$. Given that for $F^L(\nu) = 0$ this is the only way a scalar operator can appear in the OPE decomposition, we conclude as anticipated that the only allowed dimension is $d$, corresponding to a marginal operator.\footnote{Note that for $d=1$ the pole appears in the opposite half-plane compared to where we are closing the contour, namely in the lower half-plane for $G^{(1)}_{\frac{d}{2}-i\nu,MN}$ and in the upper half-plane for $G^{(1)}_{\frac{d}{2}+i\nu,MN}$. The correct prescription to continue analytically in the dimension is to keep the contribution of these poles as they cross the contour.} 

Note that the square of the boundary OPE coefficient is fixed by
$F^\perp(\pm i(\frac{d}{2}-1))$,
which is therefore required to be non-negative in a unitary theory.
The presence of the scalar operator of dimension $d$ can be avoided if the function $F^\perp(\nu)$ has a zero at $\nu = \pm i(\frac{d}{2}-1)$. Therefore we see that the boundary conditions that preserve the bulk symmetry must present this zero, while those that break the bulk symmetry do not have this zero and include the tilt operator in the spectrum. In agreement with this statement, it was observed in \cite{Ankur:2023lum} that, when the  current is coupled to a dynamical gauge field, the condition $F^\perp(\pm i(\frac{d}{2}-1)) = 0$ is necessary to keep the gauge field massless at loop level.
\end{document}